\documentstyle[twoside,fleqn,espcrc2]{article}

\makeatother

\newcommand{\ot}{\frac{1}{2}}

\title
{Lee-Yang Zeroes and Logarithmic Corrections in
the $\Phi^4_4$ Theory\thanks{Presented by R. Kenna.
Supported by Fonds zur F\"or\-der\-ung der
Wissenschaftlichen Forschung in \"Oster\-reich, project P7849.} }
\author
{R. Kenna and C.B. Lang\\~ \\
Institut f\"ur Theoretische Physik,\\
Universit\"at Graz, A-8010 Graz, AUSTRIA}

\begin{document}

\begin{abstract}
The leading mean-field critical behaviour of $\phi^4_4$-theory is
modified by multiplicative logarithmic corrections. We analyse these
corrections both analytically and numerically. In particular we present
a finite-size scaling theory for the Lee-Yang zeroes and temperature
zeroes, both of which exhibit logarithmic corrections. On lattices from
size $8^4$ to $24^4$, Monte-Carlo cluster methods and multi-histogram
techniques are used to determine the partition function zeroes closest
to the critical point. Finite-size scaling behaviour is verified and the
logarithmic corrections are found to be in good agreement with our
analytical predictions.
\end{abstract}

\maketitle

\section{INTRODUCTION}

The single component version of $\phi^4$ theory in the $d$-dimensional
Euclidean space-time continuum is defined  by the Hamiltonian density
\begin{equation}
  {\cal{H}} = \frac{1}{2}(\nabla \phi)^2 +
  \frac{m_0^2}{2}\phi^2 + \frac{g_0}{4!}\phi^4
  -H(x)\phi(x)
\label{BLZ6.36}
\end{equation}
where $H(x)$ is the source for the fields $\phi (x)$. The lattice
parameterization of the theory (in the absence of a source field)
is given by the action
\begin{equation}
  -\kappa \sum_{x,\mu}\phi_x \phi_{x+\mu}   +   \sum_x \phi_x^2
 + \lambda \sum_x\left( \phi_x^2-1  \right)^2 .
\label{Lang2.6}
\end{equation}
Here the hopping parameter $\kappa$ and the quartic coefficient
$\lambda$ correspond to the bare mass $m_0$ and bare quartic coupling
$g_0$ respectively. The limit $\lambda \rightarrow \infty$ gives the
Ising model.

Above one dimension the discretized theory exhibits a phase transition
(of second order) near which the continuum theory can be recovered. To
remove the cutoff, it turns out that the quartic coupling has to be
taken to the infra--red fixed point (IR FP) $g_R^*$. The theory is
believed to be trivial in $d=4$ --- although this has never been rigorously
proved. This means that it is in  the universality class of the theory
of free bosonic fields. The leading (mean field) scaling behaviour is
modified by logarithmic corrections, which are linked to the triviality
of the theory \cite{ADCCaFr}. Their identification provides the primary
motivation for this work.

Logarithmic corrections to scaling in the infinite volume system have
been studied in \cite{BrLeZi76} and \cite{LW}. Here we report on results
for finite-size scaling (FSS) \cite{Br82,Ba83} of the $\phi^4$ theory
which we have extended to four dimensions. Such finite size theories
can be  tested using non-perturbative (i.e., numerical) techniques.

The usual statement of FSS is the following \cite{Ba83}: For any thermodynamic
quantity $P_L(\kappa)$, measured on a system of linear extent $L$ and near
criticality,
\begin{equation}
      \frac{P_L(\kappa)}{P_\infty (\kappa)}
            = f\left( \frac{L}{\xi_\infty(\kappa)} \right),
\label{FSS}
\end{equation}
where $\xi_\infty(\kappa)$ is the correlation length of the infinite volume
system. The usual justification for this formula is that $L$ and $\xi_\infty$
are the only length scales involved and hence their ratio,
$x= L / \xi_\infty(\kappa)$, is the scaling variable. Until 1982 this
statement had the status of a hypothesis. Then, Br\'ezin \cite{Br82}
succeeded in proving (\ref{FSS}) from the renormalization group (RG).
An essential ingredient in this proof is that the running quartic
coupling be approximated by its IR FP value $g_R^*$ in the
scaling region. Now, in $d=4$ (in the perturbative formulation at least),
the IR FP  of the Callan-Symanzik beta function is at the origin.
The approximation above then leads to the mean field theory which predicts
a phase transition even for a finite system. For this reason FSS in the form
(\ref{FSS})  breaks down in $d=4$. The intuitive justification given above
is however a dimension independent argument. It is not clear, then, why it
should fail in $d=4$ while being valid for $d<4$.

We claim that the usual statement (\ref{FSS}) is, in fact, flawed and
propose a modified FSS formula, valid in any dimension including four.

\section{THE PERTURBATIVE RENORMA\-LI\-ZA\-TION GROUP}

At some critical value, ${m_0^2}_c$, of  the bare mass, the renormalized
theory is massless.  Writing $m_0^2$ as ${m_0^2}_c + t$, $t$ becomes a
measure of the deviation away from the massless theory. In the Ising
version of the model, it is proportional to $\kappa - \kappa_c$, $\kappa_c$
being the critical hopping parameter. The generating functional $W[H,t]$ is
defined by
\begin{equation}
  e^{W[H,t]} = C \int \prod_x d \phi (x) e^{-\int
d^dx{\cal{H}}},
\label{R6.1}
\end{equation}
$C$ being a normalization constant. The function conjugate to $H(x)$
is
\begin{equation}
 M(x,t) = \frac{\delta W [H,t]}{\delta H(x)} = \langle \phi(x)
\rangle_{H,t}.
\label{BLZ2.15prime}
\end{equation}
If $H$ is independent of $x$, (which we henceforth assume),
then $W$ is a function of its arguments.

The generating functional $\Gamma [M,t]$ of the one particle irreducible
vertex functions is defined through the Legendre transformation
\begin{equation}
\Gamma[M,t] + W[H,t] = \int dx H(x)M(x),
\label{Legendre}
\end{equation}
with
\begin{equation}
 H(x,t) = \frac{\delta \Gamma [M,t]}{\delta M(x)}.
\label{BLZ2.19}
\end{equation}

{\sloppy
After isolating the divergences occurring in the Schwinger functions,
one can write down the relationship between the bare and renormalized theories.
In order to be able to study the onset of criticality, in both the symmetric
and the broken phases, one first considers the massless renormalized theory
--- renormalized at some arbitrary mass-scale parameter $\mu$. Expanding in the
reduced mass $t$ and in the conjugate function $M$, gives the renormalization
group equation (RGE) for the massive theory in the critical region. Because
of the local nature of the renormalization group, the renormalization
constants of the infinite volume theory render the finite volume theory
finite too \cite{Br82}.}

For a system of finite volume $L^d$, with reduced temperature $t$ and
magnetization $M$, the above generating functional becomes the function
\[
 \Gamma (t,M,g_R,\mu,L)
\]
in which $g_R$ represents the renormalized quartic coupling. The RGE
expresses the invariance of the physics under a rescaling of $\mu$.
I.e., when the mass-scale $\mu$ is  varied, $t$,$M$ and $g_R$ respond in
a way which is governed by the flow equations \cite{BrLeZi76,LW}. In
four dimensions these flow equations can be solved perturbatively in $g_R$.
Rescaling $\mu$ to $\mu / L $, and using dimensional analysis, gives
the following solution of the RGE \cite{KeLa92}:
\begin{eqnarray}
\lefteqn{  \Gamma \left(t,M,g_R,1,L\right)
   \simeq }
    \nonumber  \\
& &  L^{-4}
  \Gamma \left(L^2t\left( \frac{2}{3g_R \ln{L}}  \right)^{\frac{1}{3}},
                   LM,\frac{2}{3 g_R \ln{L}},1,1\right)
     \nonumber \\
& &  + \frac{3}{4}
    \left( \frac{2}{3g_R}  \right)^{\frac{2}{3}}
     t^{2}
   \left( \ln{ L}  \right)^{\frac{1}{3}}.
\label{Hl202}
\end{eqnarray}
To determine how the running coupling on the right hand side of (\ref{Hl202})
couples to the remaining terms, perturbation theory must be applied to $\Gamma$
itself. This gives \cite{BrLeZi76,KeLa92}
\begin{eqnarray}
\lefteqn{
         \Gamma \left( t,M,g_R,1,L  \right)
 = c_1 tM^2 \left( \ln{L}  \right)^{-\frac{1}{3}}
}
\nonumber \\
& &
   +
   c_2 M^4 (\ln{L})^{-1}
   +
   c_3 t^2 (\ln{L})^{\frac{1}{3}}
\label{Gamma00bkn}
\end{eqnarray}
where $c_1, \dots, c_3$ are constants. Applying (\ref{BLZ2.19}) to this yields
for the external field
\begin{eqnarray}
\lefteqn{  H\left( t,M,g_R,1,L  \right)
 \simeq  }
\nonumber \\
& &
 c_4 t M (\ln{L})^{-{\frac{1}{3}}}
 +
 c_5 M^3 (\ln{L})^{-1},
\label{Hl300}
\end{eqnarray}
where, again, $c_4$ and $c_5$ are constants.

These give for the free energy per unit volume in the presence of an
external field
\begin{eqnarray}
\lefteqn{ W_L(t,H) = c_1^\prime t M^2 (\ln{L})^{-\frac{1}{3}}}
\nonumber \\
& &
   +
   c_2^\prime M^4 (\ln{L})^{-1}
   +
   c_3 t^2 (\ln{L})^{\frac{1}{3}},
\label{WtHl4D}
\end{eqnarray}
$c_1^\prime$ and $c_2^\prime$ being constants and $M$ given by (\ref{Hl300}).

If $H$ vanishes, then (\ref{Hl300}) and (\ref{WtHl4D}) give
\begin{equation}
  W_L(t,0) \propto t^2 \left( \ln{L}  \right)^{\frac{1}{3}}.
\label{Wt0l4D}
\end{equation}

One could proceed directly from (\ref{WtHl4D}) or (\ref{Wt0l4D}) to find the
FSS formulae for thermodynamic observables. But it is more complete to study
the partition function itself. This is entirely equivalent to the study of
its zeroes. For fixed real $t$ the zeroes in the complex $h$ plane are called
Lee--Yang zeroes \cite{LeYa52}, and in the absence of an external field,
the zeroes in $t$ are called Fisher zeroes \cite{Fi64}. Their FSS properties
below  four dimensions was studied in \cite{ItPeZu}. In this section, the
corresponding FSS theory is presented for four dimensions where logarithmic
corrections are manifest.

The total free energy  at the critical temperature in four dimensions in the
presence of an external field is given by (\ref{WtHl4D}) as
\begin{equation}
   L^4  (\ln{L})^{\frac{1}{3}} H^{\frac{4}{3}}.
\end{equation}
The partition function is therefore
\begin{equation}
 Z_L(t=0,H) = Q\left(    L^4  (\ln{L})^{\frac{1}{3}} H^{\frac{4}{3}}  \right).
\end{equation}
When the partition function is zero, solving for $H$ gives
\begin{equation}
 H_j \propto L^{-3} (\ln{L})^{-\frac{1}{4}}
\label{LYzeroesd=4}
\end{equation}
where the constant of proportionality depends on the index $j$ of the zero.
This is the FSS formula for Lee--Yang zeroes in four dimensions.

If $H$ vanishes, (\ref{Wt0l4D}) can be used in a similar way to show that the
Fisher zeroes scale as
\begin{equation}
 t_j \propto  L^{-2} \left( \ln{L} \right)^{-\frac{1}{6}}.
\label{fssfisher4D}
\end{equation}

Once the FSS behaviour of the partition function zeroes has been found one
can easily find the corresponding behaviour for thermodynamic functions by
expressing them in terms of the zeroes. These considerations give for the
zero field magnetic susceptibility and specific heat
\begin{equation}
 \chi_L \left( t=0,H=0 \right)
 \propto
 L^2 \left( \ln{L} \right)^{\frac{1}{2}}
\end{equation}
and
\begin{equation}
 C_L\left(t=0,H=0\right)
 \propto
 \left( \ln{L} \right)^{\frac{1}{3}}.
\label{Cl4D}
\end{equation}

\section{NON-PERTURBATIVE ANALYSIS OF FINITE SIZE SCALING}
\setcounter{equation}{0}

The Swendsen--Wang cluster algorithm \cite{SwWa87} was applied to the
Ising version of the theory on lattices of sizes $8^4$ to $24^4$.

In an external field $h$ ($=\kappa H$), the partition function can be
written as
\begin{equation}
Z(\kappa,h) = \sum_{M=-N}^{N} \sum_{S=-4N}^{4N}
  \rho(S,M) e^{\kappa S + hM},
\label{Znum}
\end{equation}
where
\begin{equation}
 S = \sum_x \sum_{\mu=1}^{4} \phi_x \phi_{x+\mu}
\quad , \quad  M = \sum_x \phi_x, \quad
\end{equation}
and the spectral density $\rho(S,M)$ is the relative weight of configurations
having given values of $S$ and $M$. The `multihistogram' method \cite{FeSw88}
was used to combine histograms determined at various values of $\kappa$. This
provides an optimal estimator for the spectral density and allows one to
construct $Z(\kappa,h)$ in the complex neighbourhood of the critical point.
A Newton--Raphson algorithm was used to determine nearby zeroes.

The leading (power law) FSS behaviour of the Lee--Yang and Fisher zeroes
was found to be slightly deviant from the mean field predictions. These
deviations find their explanation in the presence of logarithmic corrections.
To isolate these corrections, in the case of Fisher zeroes, we plot
in fig.1a $\ln{(L^2 {\rm{Im}} \kappa_1)}$ versus $\ln{(\ln{L})}$. The
negative slope is in good agreement with the scaling prediction of
$- \frac{1}{6}$. In fact, a fit to all five points gives a slope $-0.217(12)$.
Excluding the point corresponding to $L = 8$ gives a slope of $-0.21(4)$. The
solid line is the best fit to the remaining points  assuming the
theoretical prediction $-\frac{1}{6}$ from (\ref{fssfisher4D}).

We may now determine $\kappa_c$ from
$\left| \kappa_j - \kappa_c\right|  \propto
l^{-2} \left( \ln{l} \right)^{- 1/6}$.
Using the first Fisher zeroes, we find $\kappa_c \simeq 0.149703(15)$
in good agreement with the value $0.149668(30)$ from high temperature
expansions \cite{GaSyMc79}.

To identify the logarithmic corrections for the Lee--Yang zeroes, we plot
in fig.1b $\ln{( L^3 {\rm{Im}} h_1)}$ against $\ln{(\ln{L})}$.
A best fit to all five points gives a slope of $-0.204(9)$ which compares
well with the theoretical prediction of $-\frac{1}{4}$ from
(\ref{LYzeroesd=4}). Excluding the smallest lattice, a fit to the
remaining four points gives a slope $-0.22(3)$. The solid line in  fig.1b
is the best fit to the last four points with given slope $-\frac{1}{4}$.

\begin{figure}[htb]
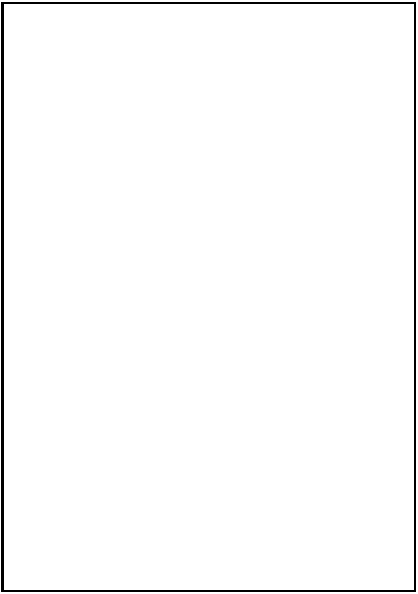

\framebox[55mm]{\rule[-21mm]{0mm}{76mm}}
\caption{Logarithmic corrections to FSS of (a) Fisher zeroes
and (b) Lee--Yang zeroes.}
\label{fig:largenenough}
\end{figure}

\section{CONCLUSIONS}
\setcounter{equation}{0}

A finite size scaling theory has been developed for the single component
$\phi^4$ theory in $d=4$ dimensions. Emphasis has been placed on logarithmic
corrections to the mean field predictions. This has been checked
non-perturbatively using high precision numerical methods, and good agreement
is found.

FSS formulae for other thermodynamic functions are also given. These exhibit
logarithmic corrections too.  The FSS formula for the correlation length of
a four dimensional system also involves logarithmic corrections. This was
derived by Br\'ezin \cite{Br82} for a system of extent $L$ in all directions.
At the infinite volume critical point
$\kappa = \kappa_c$,
\begin{equation}
      \xi_L(\kappa_c ) \propto L (\ln{L})^{\frac{1}{4}}.
\end{equation}
This suggests that the  FSS variable should be
\begin{equation}
\frac{\xi_L(\kappa_c)}{\xi_\infty(\kappa)} =
\frac{L (\ln{L})^{\frac{1}{4}}  }{t^{-\ot} \mid\;\ln t\mid\;^{\frac{1}{6}}}
\end{equation}
in four dimensions\cite{KeLa91}.
Indeed, replacing the scaling variable, $x$. of the right hand side of
(\ref{FSS}) by the ratio $\xi_L(\kappa_c) / \xi_\infty(\kappa) $ is
sufficient to recover all the FSS formulae presented here while still
being correct in $d<4$ dimensions. We suggest that this modified FSS
hypothesis is the more appropriate one.


\end{document}